\documentclass[a4paper,twoside,reqno]{bjp}
\usepackage{graphicx}
\usepackage{cite}
\usepackage{amssymb,amsmath,amscd,amsthm}
\usepackage{times}

\usepackage[bookmarks=false]{hyperref}
\hypersetup{%
    colorlinks=true,        
    linkcolor=blue,          
    citecolor=blue,         
    urlcolor=blue           
    }

\usepackage{geometry}
\allowdisplaybreaks

\begin{document}

\title{Shape coexistence in $^{74}$Ge, $^{74}$Se and $^{74}$Kr investigated by phenomenological and microscopic models}

\runningheads{P. Buganu, \textit{et al.}}{Shape coexistence investigated by phenomenological and microscopic models}

\begin{start}{%
		
\author{A. Ait Ben Mennana}{1},
\author{R. Benjedi}{1},	
\author{P. Buganu}{2},
\author{R. Budaca}{2},\\
\author{A. I. Budaca}{2},
\author{Y. EL Bassem}{1,3},
\author{A. Lahbas}{3,4},
\author{M. Oulne}{1}

\address{LPHEA, Department of Physics, Faculty of Sciences Semlalia, Cadi Ayyad University, Marrakesh, Morocco}{1}
\address{Department of Theoretical Physics, Horia Hulubei National Institute for R and D in Physics and Nuclear Engineering, Str. Reactorului 30, RO-077125, POB-MG6, Bucharest-M\v{a}gurele, Romania}{2}
\address{ERMAM, Polydisciplinary Faculty of Quarzazate, Ibn Zhor University, Quarzazate Box 638, Morocco}{3}
\address{ESMaR, Department of Physics, Faculty of Sciences, Mohammed V University in Rabat, Morocco}{4}

}

\begin{Abstract}
The deformation properties of $^{74}$Ge, $^{74}$Se and $^{74}$Kr are studied within the phenomenological Bohr-Mottelson model, having as input the experimental collective energy states, as well with Covariant Density Functional theories based on microscopic structural information. The results of these approaches are shown to be compatible in what concerns the presence of coexisting shapes in the considered nuclei, while the emergence of shape mixing is deduced from the phenomenological calculated collective states.
\end{Abstract}

\begin{KEY}
Bohr-Mottelson model, covariant density functional theory, the low-lying states, shape coexistence and mixing
\end{KEY}
\end{start}


\section{Introduction}

\par Shape coexistence, being a subject of large interest in nuclear physics, is a very peculiar nuclear phenomenon when two or more states with widely different deformation properties occur in the same nuclei within a very narrow energy range at low excitation energy  \cite{Poves}. The presence of the second ${0}^{+}$ state as the first excited state in even-even nuclei, is one of the most important signatures that indicates the shape coexistence \cite{Heyde1,Andreyev}. 
In the present study, the phenomenon will be addressed in the frame of both phenomenological and microscopic models. The first, corresponding to the Bohr-Mottelson model (BMM)\cite{Bohr1,Bohr2}, describes the low-lying spectra of nuclei in terms of vibrations and rotations of their ground state shape. It's an appropriate model to investigate shape coexistence and mixing phenomena \cite{Heyde1}. The second model corresponds to the Covariant Density Functional theories (CDFT)  \cite{Dirac,Negele,Niksic1,Lalazissis,Roca,Niksic2}.
A sixth order anharmonic oscillator potential (sextic) is used for the Bohr Mottelson Hamiltonian, being numerically diagonalized \cite{Budaca1} in a basis of Bessel functions of the first kind \cite{Taseli1}, which in turn are solutions for the infinite square well potential \cite{Iachello1,Iachello2}. This method has been already involved with success to describe shape coexistence phenomenon within some nuclei as $^{76}$Kr \cite{Budaca2}, $^{72,74,76}$Se \cite{Budaca3} and $^{96,98,100}$Mo \cite{Budaca4}. 
Additionally, it can be used to investigate the shape mixing and coexistence phenomena by increasing the height of the barrier \cite{Budaca2,Budaca5}. The free parameters of the phenomenological model are determined by fitting the experimental data and then are used to calculate the energy spectrum of the ground, $\beta$ and $\gamma$ bands, the quadrupole $(E2)$ and monopole $(E0)$ transitions, respectively the shape of the ground and excited states. Concerning the microscopic methods, based on the CDFT, are involved in obtaining the quadrupole deformation by calculating and analyzing the potential energy surfaces (PES). CDFT is one of the most attractive form of the nuclear	density functional theories, proving to be very successful in the nuclear structure analysis, as well as in describing the ground and excited states throughout the nucleic chart \cite{Bassem1,Bassem2,Agbemava}. Applications with these models are done for three isobars with mass number $A=74$, namely $^{74}$Ge $(Z=32)$, $ ^{74}$Se $(Z=34)$ and $ ^{74}$Kr $(Z=36)$, which, based on previous works \cite{Nomura,Bender,Budaca3}, are suspected of presenting shape coexistence phenomenon. Additionally, in $A\sim70$\cite{Benjedi}, the analysis of the evolution with a mass number of the measured energies for the  $0^{+}_2$ and $2^{+}_1$ states, and of the monopole strength connecting the lowest $0^{+}$ states, along the isotopic chains of Ge, Se, and Kr deduces that the three nuclei in $A=74$ should present shape coexistence and mixing in their ground states.

The plan of the paper is the following. In Section 2, the models and the most relevant quantities to investigate the shape coexistence phenomenon are introduced, while Section 3 is dedicated for numerical applications to experimental data of $^{74}$Ge, $^{74}$Se and $^{74}$Kr. Finally, the main achievements of the work are highlighted in Section 4.

\section{ Presentation of the models}
\renewcommand{\theequation}{2.\arabic{equation}}
In this study,  phenomenological (BMM) and microscopic (CDFT) models are used to investigate the presence of the shape coexistence phenomenon within three nuclei with atomic mass $A=74$ (Ge, Se, Kr). Before discussing the numerical applications to experimental data, which are elaborated in the next section, a brief presentation of the models will be made here.

\subsection{Bohr Hamiltonian with sextic oscillator potential}
 For $\gamma$-unstable nuclear shapes, the collective potential $V$ depends only on the $\beta$ variable, allowing  the separation of the $\beta$ variable from the $\gamma$ variable and the Euler angles \cite{Wilets,Bes}. This leads to the following $\beta$ equation:
\begin{equation}
\left[-\frac{d^{2}}{d\beta^{2}}-\frac{4}{\beta}\frac{d}{d\beta}+\frac{\tau(\tau+3)}{\beta^{2}}+v(\beta)\right]\psi(\beta)=\varepsilon \psi(\beta),
\label{eqbeta}
\end{equation}
  where $\tau$  denotes the seniority quantum number, which defines the eigenvalue of the Casimir operator of the SO(5) group \cite{Rakavy} describing the coupling of rotational degrees of freedom with the fluctuation of the $\gamma$ shape variable, while $v(\beta)=(2B/\hbar^{2})V(\beta)$ and $\varepsilon=(2B/\hbar^{2})E$ are the reduced potential and total energy.
The expression of the sextic potential is considered for Eq. (\ref{eqbeta}):
\begin{equation}
v(\beta)=\beta^{2}+v_{1}\beta^{4}+v_{2}\beta^{6},
\label{potbeta}
\end{equation}
which due to the scaling property of the problem is uniquely determined only by two free parameters, $v_{1}$ and $v_{2}$ \cite{Budaca1}. It is the simplest allowed potential which can present two minima, a spherical and a deformed one, separated by a barrier. To find the energy $\varepsilon$, Eq. (\ref{eqbeta}) with potential (\ref{potbeta}) is diagonalized in a basis of Bessel functions of the first kind \cite{Budaca4}. The basis functions are solutions of the same equation but for an infinite square well potential (ISWP) \cite{Iachello1} encompassing the relevant part of the adopted sextic potential:
\begin{equation}
f_{n,\nu}(\beta)=\frac{\sqrt{2}}{\beta_{w}}\frac{\beta^{-\frac{3}{2}}J_{n,\nu}\left(\frac{z_{n,\nu}}{\beta_{w}}\beta\right)}{J_{\nu+1}(z_{n,\nu})},\;\;\nu=\tau+\frac{3}{2}.
\end{equation}
Here, $\beta_{w}$ is the width of ISWP, while $z_{n,\nu}$ is the n$^{th}$ zero of the Bessel function $J_{\nu}(z)$. Therefore, the wave function for the sextic potential is written as an expansion in this basis:
\begin{equation}
F_{\xi,\tau}(\beta)=\sum_{n=1}^{n_{Max}}A_{n}^{\xi}f_{n,\nu(\tau)}(\beta),
\label{wavef}
\end{equation}
where $n_{Max}$ is the dimension of the truncation basis, $\xi=1,...,n_{Max}$ is related with the $\beta$ vibration quantum number $(\xi=n_{\beta}+1)$, while $A_{n}^{\xi}$ are the eigenvector components following to be determined from the diagonalization of the corresponding Hamiltonian matrix. As in \cite{Budaca4}, the Hamiltonian model is amended with an SO(5) symmetric term $\hat{L}^{2}$, which splits the  $\tau$ energy multiplet without changing the wave function \cite{Caprio}. In this way, the quantitative description of the experimental data is improved. Basically, this term gives a contribution only to the total energy:
\begin{equation}
E_{\xi,\tau,L}=\frac{\hbar^{2}}{2B}\left[{\varepsilon}_{\xi,\tau}(v_{1},v_{2})+dL(L+1)\right],
\label{enlev}
\end{equation}
where ${\varepsilon}_{\xi,\tau}(v_{1},v_{2})$ is the eigenvalue energy of diagonalization procedure. Finally, the energy depends on three free parameters $v_{1}$, $v_{2}$ and $d$, respectively on a scaling parameter $(\hbar^{2}/2B)$.
The wave function (\ref{wavef}) and the SO(5) spherical harmonics \cite{Rowe1} are then used to calculate some electromagnetic observables such as quadrupole and monopole transition probabilities. For $E2$ rate, one uses the quadrupole transition operator as in \cite{Budaca4}:
\begin{equation}
T_{\mu}^{(E2)}=\frac{3R^{2}Ze}{4\pi}\beta_{M}\beta\times  \left[D_{\mu,0}^{2}(\Omega)\cos\gamma+\frac{1}{\sqrt{2}}[D_{\mu,2}^{2}(\Omega)+D_{\mu,-2}^{2}(\Omega)]\sin\gamma\right],
\label{tranop}
\end{equation}
where $\beta_M$ is a scaling factor relating the quadrupole deformation $\beta_2$ with the $\beta$ shape variable ($\beta_2=\beta_M\beta$), $R$ is the nuclear radius, $Z$ is the charge number and $e$ the elementary charge. Whereas the monopole strength is determined as \cite{Wood2,Bonnet}:
\begin{equation}
\rho^{2}(E0;i\rightarrow f)=\left(\frac{3Z}{4\pi}\right)^{2}\beta_{M}^{4}\langle F_{i}(\beta)|\beta^{2}|F_{f}(\beta)\rangle^{2}.
\label{rho0}
\end{equation}
\subsection{Covariant Density Functional Theory (CDFT)}

The calculations are done in the framework of CDFT \cite{Niksic1,Lalazissis,Roca,Niksic2}, being used the very successful density-dependent meson-exchange DD-ME2 \cite{Lalazissis}, which has a finite interaction range and the parameter sets used in \cite{Benjedi}.\\
The standard Lagrangian density with medium dependence vertices that defines the meson-exchange model~\cite{Gambhir} is given by:
\begin{eqnarray}
\mathcal{L}&=&\bar{\psi}\left[
\gamma(i\partial-g_{\omega}\omega-g_{\rho
}\vec{\rho}\,\vec{\tau}-eA)-m-g_{\sigma}\sigma\right]  \psi\nonumber\\
&&+\frac{1}{2}(\partial\sigma)^{2}-\frac{1}{2}m_{\sigma}^{2}\sigma^{2}-\frac{1}{4}\Omega_{\mu\nu}\Omega^{\mu\nu}+\frac{1}{2}m_{\omega}^{2}\omega
^{2}\nonumber\\
&&-\frac{1}{4}{\vec{R}}_{\mu\nu}{\vec{R}}^{\mu\nu}+\frac{1}{2}m_{\rho}%
^{2}\vec{\rho}^{\,2}-\frac{1}{4}F_{\mu\nu}F^{\mu\nu},
\label{lagrangian}
\end{eqnarray}
where $m$ is the bare nucleon mass and $\psi$ denotes the Dirac spinors. $m_\rho$, $m_\sigma$ and $m_\omega$ are the masses of  $\rho$ meson,  $\sigma$ meson and  $\omega$ meson,
with the corresponding coupling constants for the mesons to the nucleons as $g_\rho$, $g_\sigma$ and $g_\omega$ respectively, and $e$ is the charge of the proton.
\section{Numerical results}
The phenomenological and microscopic models presented in Section 2 are involved to investigate the structure of the states and the corresponding shape for  $^{74}$Ge and $^{74}$Kr. The $^{74}$Se nucleus has been already treated in \cite{Budaca4} in the frame of the Bohr model, finding a coexistence between a spherical shape and a prolate one for its structure, while here the results will be only compared with those coming from microscopic calculations. In the frame of BMM, $^{74}$Ge and $^{74}$Kr are treated as being good candidates for coexistence between spherical and $\gamma$-unstable shapes. There are several signatures or arguments which support this supposition. For example, some representative energy and $B(E2)$ ratios, given in Table \ref{tab1}, indicate that the structure of $^{74}$Ge and $^{74}$Kr are close to  O(6) \cite{Wilets}, respectively E(5) \cite{Iachello1} structure. Even if the $0_{\beta}^{+}$ state of $^{74}$Kr is much lower in energy than in the case of E(5), in turn there is a good match for the $B(E2;0_{\beta}^{+}\rightarrow2_{g}^{+})$ transition.

\begin{table}[h]
	\centering
	\caption{Signatures of $\gamma$-unstable structure in $^{74}$Ge and $^{74}$Kr.}
	\resizebox{7.5cm}{1.5cm} {
		\begin{tabular}{ccccc}
			\hline
			\hline
			$\frac{\mathrm{Models}}{\mathrm{Nuclei}}$&$\frac{E(4_{g}^{+})}{E(2_{g}^{+})}$&$\frac{E(0_{\beta}^{+})}{E(2_{g}^{+})}$&$\frac{B(E2;4_{g}^{+}\rightarrow2_{g}^{+})}{B(E2;2_{g}^{+}\rightarrow0_{g}^{+})}$&$\frac{B(E2;0_{\beta}^{+}\rightarrow2_{g}^{+})}{B(E2;2_{g}^{+}\rightarrow0_{g}^{+})}$\\
			\hline
			O(6)\cite{Wilets}&2.50&$-$&1.43&$-$\\
			$^{74}$Ge\cite{Balraj}&2.46&2.49&1.24&0.27\\
			\hline
			E(5)\cite{Iachello1}&2.20&3.03&1.68&0.87\\
			$^{74}$Kr\cite{Balraj}&2.22&1.12&2.34&0.90\\
			\hline
			\hline
	\end{tabular}}
	\label{tab1}
\end{table}

\begin{figure}[h]
	\centering
	\includegraphics[width=0.9\textwidth]{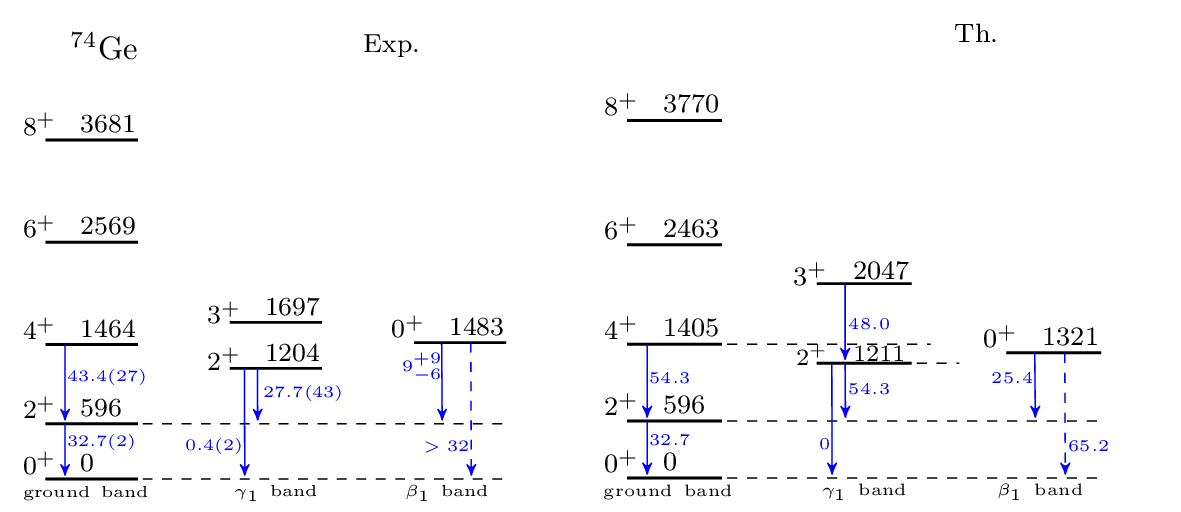}\\
	\caption{(Color online) Energy levels (\ref{enlev}) in keV, $B(E2)$ transitions probabilities in W.u. (full vertical arrows) and monopole transition $\rho^2(E0)$ (\ref{rho0}) (dashed vertical arrow) are compared with the experimental data \cite{Balraj} for $^{74}$Ge.}
	\label{fig1}
\end{figure}

\begin{figure}[h]
	\centering
	\includegraphics[width=0.9\textwidth]{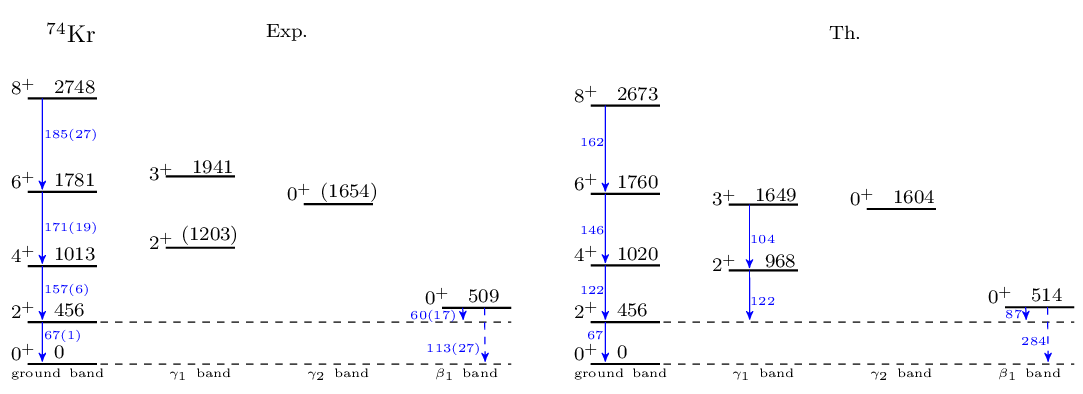}\\
	\caption{(Color online) The same as in Fig.\ref{fig1}, but for experimental data \cite{Balraj} of $^{74}$Kr. }
	\label{fig2}
\end{figure}

The energy levels, the $(E2)$ transition probabilities and monopole $(E0)$ transition obtained for $^{74}$Ge and $^{74}$Kr, are compared in Figs. \ref{fig1} and \ref{fig2} with experimental data. In Fig. \ref{fig1}, corresponding to  $^{74}$Ge, the energy levels of the ground band, first excited $0^{+}$, respectively $2^{+}$ and $3^{+}$ of the $\gamma$ band, as well as the few available experimental $B(E2)$ transitions and the monopole transition, are very well reproduced. Concerning $^{74}$Kr, which is shown in Fig. \ref{fig2}, one finds a similar situation, with a good description of the ground band, of the first excited $0^{+}$, respectively of some states of the $\gamma$ band. Additionally here, one has an experimental data for a second excited $0^+$, which is almost degenerated with $6^{+}$ of the ground band, respectively $3^{+}$ of the $\gamma$ band. These three states are very well described by the $\tau=3$ multiplet, which in the present model is slightly splitted over the total angular momentum such that to obtain also a quantitative agreement with the experimental data.  Another important remark is that the first excited $0^{+}$ is very low in energy, especially for $^{74}$Kr, while the monopole transition between this state and ground state has large value, which could be an indication for the presence of the shape mixing and coexistence phenomena \cite{Wood2,Budaca2}. Nevertheless, the phenomological model succeeds to describe such a lower excited $0^{+}$ state, as well as the monopole transition $\rho^{2}(E0;0_{2,0}^{+}\rightarrow0_{1,0}^{+})$ even if for the latter one is used the value of $\beta_{M}$ fixed for the $B(E2;2_{1,1}^{+}\rightarrow0_{1,0}^{+})$ transition.

\begin{figure}[h]
	\centering
	\includegraphics[width=0.9\textwidth]{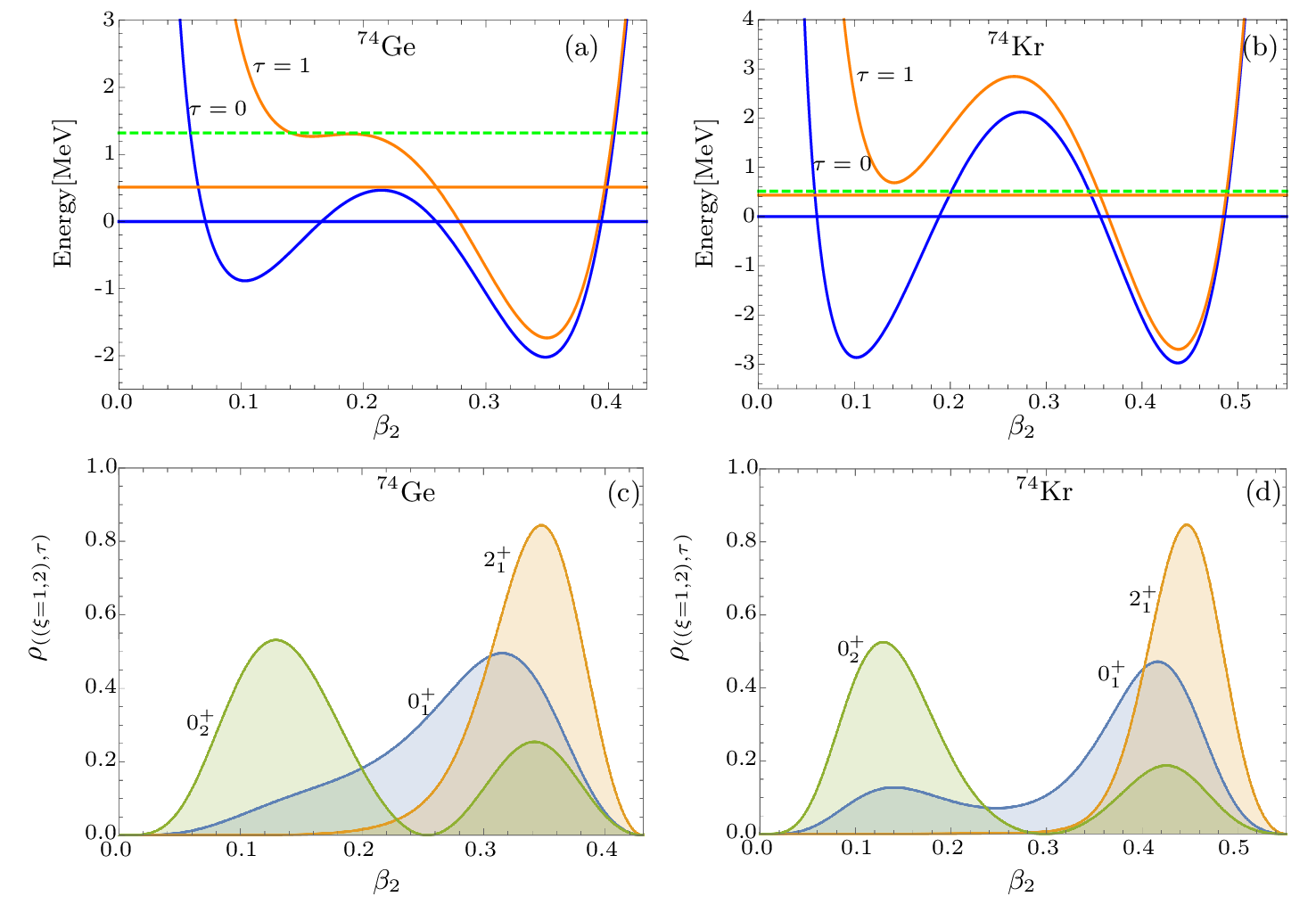}\\
	  \caption{(Color online) Effective potential (\ref{effpot}) for $\tau=0$ and $\tau=1$, energy levels (\ref{enlev}) for $(\xi=1;\tau=0,1.)$ (full horizontal lines), respectively $(\xi=2;\tau=0.)$ (dashed horizontal lines), as a function of the quadrupole deformation $\beta_{2}$ are presented in panel (a) for $^{74}$Ge, respectively (b) for $^{74}$Kr. The probability density distribution for deformation (\ref{densitybeta}) as a function of $\beta_{2}$, for the same states, is shown in panels (c) and (d) for $^{74}$Ge and $^{74}$Kr respectively. For a better correspondence, in panels (a) and (b), the energy and the effective potential of the same $\tau$ are represented by the same color. }
	\label{fig3}
\end{figure}

In order to answer the question, whether or not the shape coexistence is present in these nuclei, besides the above discussion around the lower excited $0^+$ state and large monopole transition, further, one makes a discussion on the plots for probability density distribution  of deformation (Fig. \ref{fig3}) in correlation with some representative energy levels and their corresponding effective energy potential:
\begin{equation}
V_{eff}(\beta)=\frac{\hbar^{2}}{2B}\left[\frac{\tau(\tau+3)+2}{\beta^{2}}+\beta^{2}+v_{1}\beta^{4}+v_{2}\beta^{6}\right],
\label{effpot}
\end{equation}
\begin{equation}
\rho_{\xi,\tau}(\beta)=[F_{\xi,\tau}(\beta)]^{2}\beta^{4}.
\label{densitybeta}
\end{equation}

Therefore, in Fig. \ref{fig3}, are plotted effective potentials (\ref{effpot}), absolute energies (\ref{enlev}) and probability density distribution of deformation (\ref{densitybeta}) as a function of the quadrupole deformation $\beta_{2}$ for $^{74}$Ge $(a)$ and $^{74}$Kr $(b)$ nuclei. For $^{74}$Ge, the small deformation minimum is higher than the high deformation one, while in the $^{74}$Kr case, the two minima are almost degenerated. The barrier separating the two potential minima in the ground state is smaller for the $^{74}$Ge nucleus, but is above the corresponding energy level for both nuclei. This implies that for $^{74}$Ge there is a higher amount of tunneling between the two minima and consequently a strong mixing between the two deformation configurations which are also closer to one another. The plot of the density of deformation probability for the $^{74}$Ge ground state (Fig. \ref{fig3}(c)) confirms this behaviour. It presents a very extended shape enveloping clearly distinct peaks centered on the two potential minima, with an enhanced probability above the large deformation potential minimum. The large superposition between the two peaks is due to the closeness of the two minima and their large mixing. The same mechanism acts in the ground state of $^{74}$Kr, where the two potential minima are however more displaced from each other. The mixing of two configurations provides this time well separated deformation probability peaks localized in the corresponding potential minima (Fig. \ref{fig3}(d)). 
Once again, the high deformation configuration is favored. The two-peak distribution of the deformation probability in the ground state of both nuclei is a proof of shape coexistence with mixing. It also explains the observed high monopole strength connecting ground state with the first excited $0^{+}$ state.
 Because the ground state acquires two peaks, its overlap with the wave function corresponding to the beta excited $0^{+}$ state, drastically increases because the later also have a two-peak structure as a consequence of its vibrational nature (see Fig. \ref{fig3}(c) and \ref{fig3}(d)). The effective potential associated ($\tau=1$) in the ground state for both nuclei, the $^{74}$Ge nucleus loses its double minimum shape while the $^{74}$Kr nucleus keeps it. From Fig. \ref{fig3}, it is clear that the rotationally excited state ($\tau=1$) for both nuclei has a single peak structure almost completely localized in the highly deformed potential minimum. 

\begin{figure*}
	\begin{tabular}{lrc}
		\includegraphics[width=.34\textwidth]{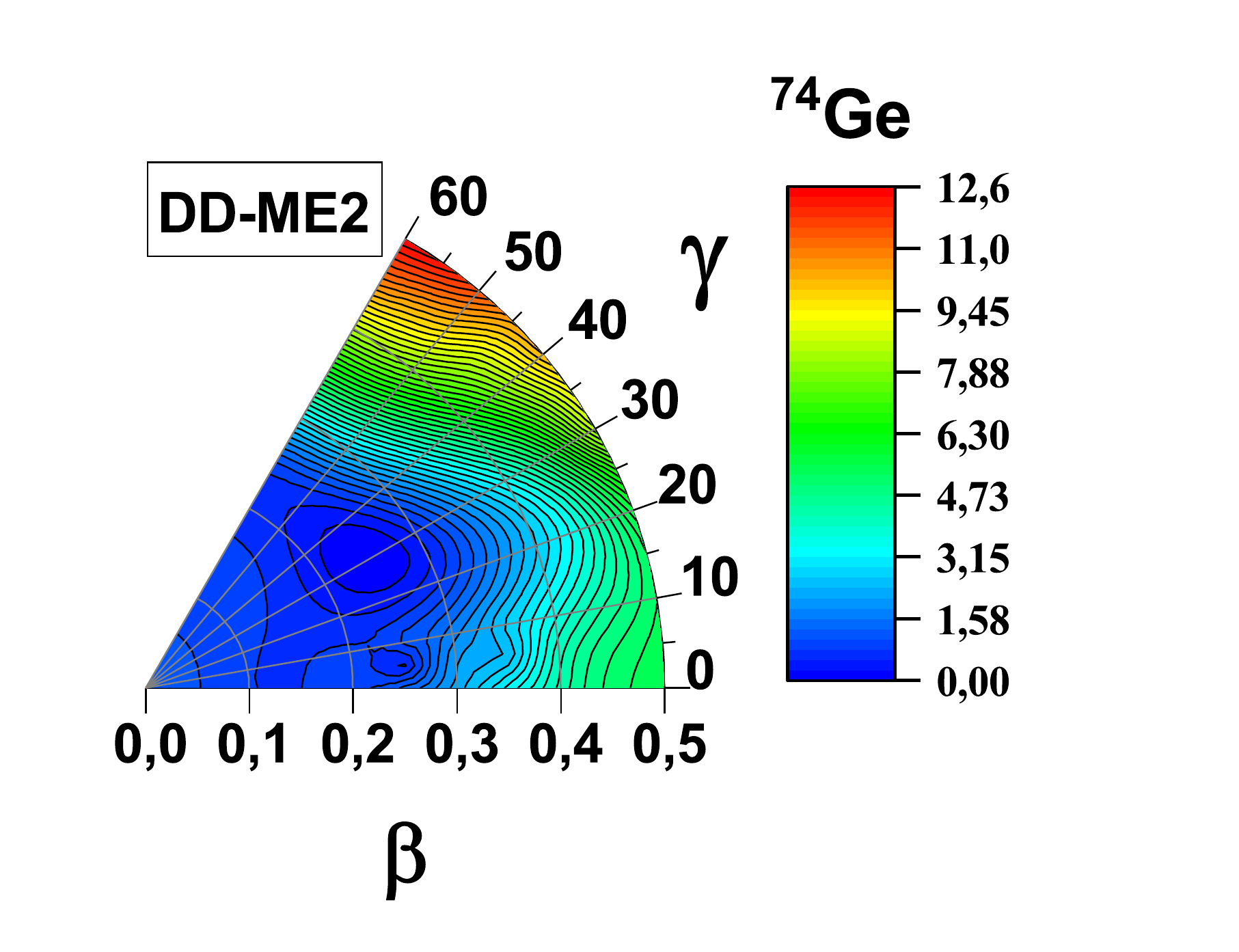}
		\includegraphics[width=.34\textwidth]{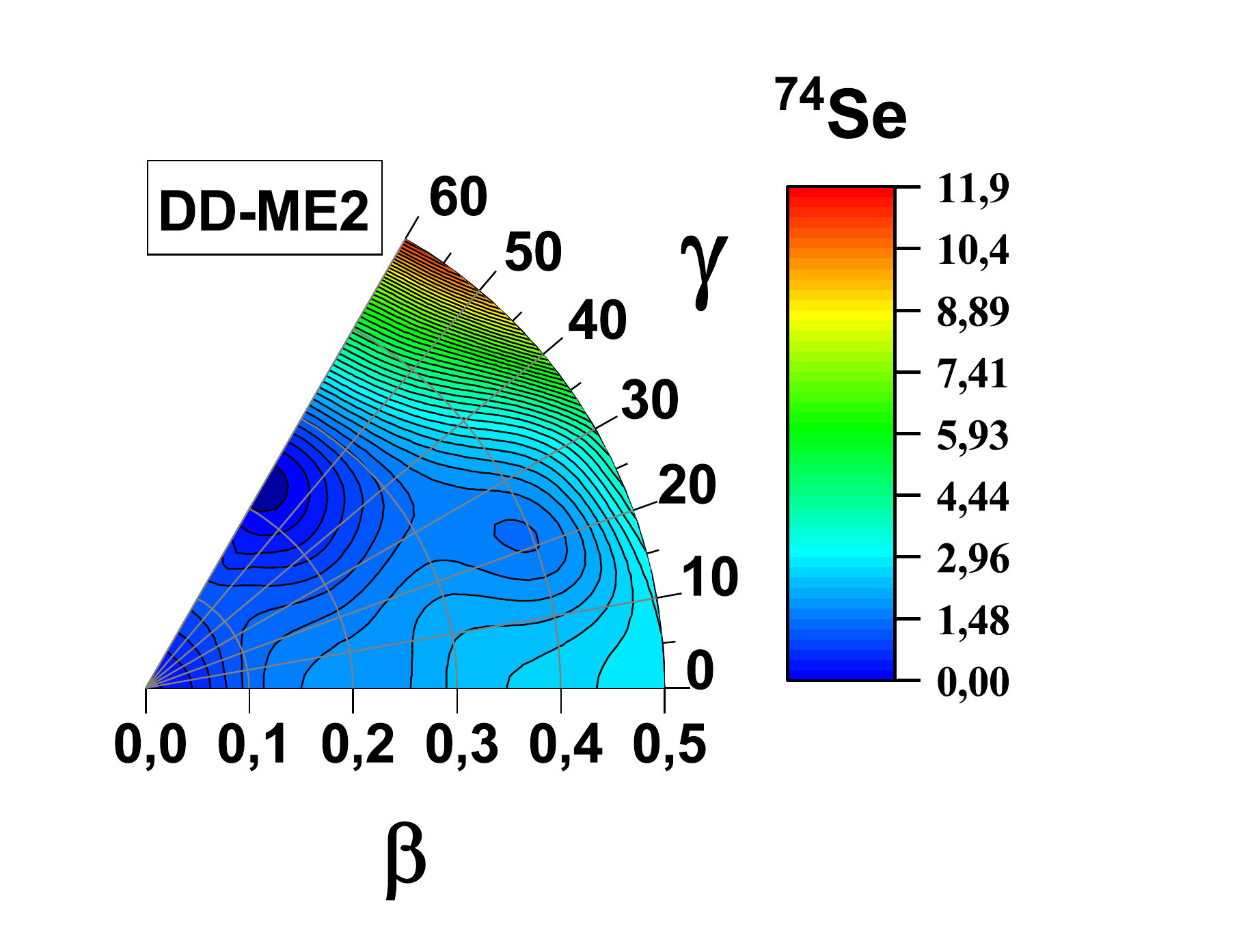}
		\includegraphics[width=.34\textwidth]{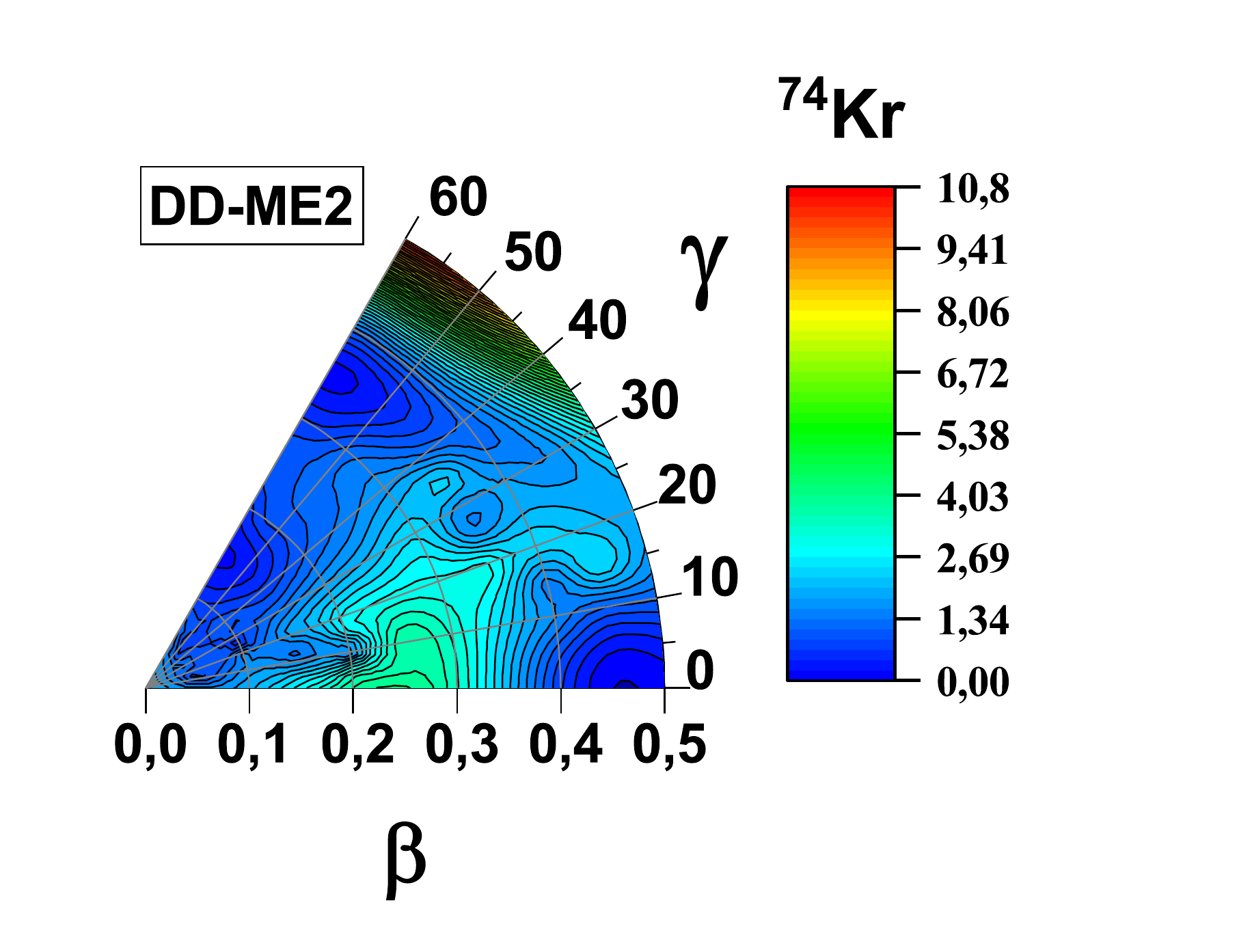} 
	\end{tabular}
	\caption{(Color online) Potential energy surfaces in the ($\beta$, $\gamma$) plane for $^{74}$Ge, $^{74}$Se and $^{74}$Kr, obtained from a CDFT calculations with the DD-ME2 parameter set. The color scale shown at the right is in MeV units and scaled such that the ground state energy to be zero.}
	\label{fig4}
\end{figure*}

Because the identification of shape coexistence phenomenon represents a challenging task when its presence has to be theoretically established within a nucleus, in the present work, we also involved in calculations one microscopic approach namely CDFT. This last approach has been used to calculate the potential energy surfaces (PES) for $^{74}$Ge (Z=32), $^{74}$Se (Z=34) and $^{74}$Kr (Z=36) in the ($\beta$, $\gamma$) plane in order to study the deformation of the ground state, and the possible shape coexistence. This is done systematically within the constrained triaxial calculations mapping the quadrupole deformation space defined by $\beta_2$ and $\gamma$ using DD-ME2 parameterizations. The obtained results are shown in Fig. \ref{fig4}.
For $^{74}$Ge, one observes a coexistence between a spherical ground state and a triaxial minimum located at $\beta_2 = 0.25$ and $\gamma = 30^{\circ}$.The same result has been found in Ref. \cite{Nomura} with the Gogny-D1M EDF. These results support the treatment of $^{74}$Ge, in the frame of the Bohr Hamiltonian with sextic potential, as manifesting a coexistence between spherical and $\gamma$-unstable shapes. Actually, by $\gamma$-unstable case one understands that the distribution of the wave function in the $\gamma$-variable has a flat peak centered around $\gamma=30^{\circ}$, covering the all range between $\gamma=0^{\circ}$ (prolate) and $\gamma=60^{\circ}$ (oblate).
Also, a coexistence between spherical and oblate ($\beta_2 = 0.2$) configurations has been obtained for $^{74}$Se in DD-ME2. Our results, in this parameterization, agree with those obtained within the five-dimensional (5D) collective Hamiltonian approach based on the Gogny-D1S energy density functional (EDF) \cite{Gaudefroy} and with the Gogny-D1M EDF \cite{Nomura}. These findings support the assumption made in \cite{Budaca3}, that the structure of $^{74}$Se is very well explained if one takes into account a presence of a coexistence between spherical and axial symmetric shapes in the frame of the Bohr Hamiltonian with sextic potential.
Concerning $^{74}$Kr, PES shows an unusual shape coexistence phenomena in the ground state. One has a very complex structure which shows up with three different minima, all of them being of axial nature, two oblate and one prolate (Fig. \ref{fig4}). The most pronounced oblate minimum is located at $\beta_2 = 0.15$ and the other has $\beta_2 = 0.35$, while the prolate minimum is located at $\beta_2 = 0.45$. These results are in agreement with those obtained in Ref. \cite{Abusara,Bender}. On the other hand, if one takes into account excited states, as in BMM, the structure of $^{74}$Kr seems to be more appropriately described by a coexistence between spherical and a $\gamma$-unstable shapes, which apparently do not fit the microscopic predictions for the ground state. An agreement between phenomenological and microscopic models can be reached if one approximates the most pronounced oblate minimum ($\beta_2 =0.15$) with a spherical one, while the other two more deformed minima, oblate ($\beta_2 =0.35$) and prolate ($\beta_2 =0.45$) ones, as a $\gamma$ "oscillation" (unstable) between prolate and oblate limits. Of course that, another possibility would be to treat $^{74}$Kr, in the frame of the Bohr Hamiltonian with sextic potential, as in the case of $^{76}$Kr \cite{Budaca2}, as having a coexistence between an approximately spherical shape  and a prolate one, but this would not reproduce well the staggering of the $\gamma$ band and the approximate degeneracy between the ground band states and those of the $\gamma$ band  \cite{Benjedi}. One can remark that the obtained results here, coming also to support the assumptions made with the Bohr Hamiltonian with sextic potential for these nuclei.
\section{Conclusions}
In conclusion, the shape coexistence phenomenon has been investigated for three nuclei with atomic mass $A=74$, namely $^{74}$Ge, $^{74}$Se and $^{74}$Kr using both phenomenological and microscopic models as the BMM and CDFT respectively. By analyzing all the data offered by these models in respect to the corresponding experimental data, it has outlined a presence of the shape coexistence in these nuclei and the fact that the most appropriate interpretation of the results would be that one has a coexistence between spherical and $\gamma$-unstable structures for $^{74}$Ge and $^{74}$Kr, respectively, a coexistence between spherical and axial symmetric structures for $^{74}$Se. Some further investigations are necessary to be done for $^{74}$Kr, where the PES obtained with CDFT indicates a presence of three minima, two oblate and a prolate one, while the Bohr model calculations suggest a very well defined shape coexistence in the ground state comparing to the other two nuclei.
\section*{Acknowledgements} 
This work was supported by High Energy Physics and Astrophysics Laboratory, Faculty of Science Semlalia, Cadi Ayyad University-Marrakesh, Morocco and by Project PN-19-06-01-01/2019 of Romanian Ministry of research, innovation and digitalization.


\end{document}